\documentclass[twocolumn]{aastex631}

\usepackage{comment}

\usepackage{amsmath}
\usepackage{multirow}

\begin{document}

\title{A geometric approach to estimate background in astronomical images}

\author[0009-0009-7497-3431]{Pushpak Pandey}
\affiliation{Inter-University Centre for Astronomy and Astrophysics, Ganeshkhind, Post Bag 4, Pune 411007, India}
\email{pushpak@iucaa.in}

\author[0000-0002-8768-9298]{Kanak Saha}
\affiliation{Inter-University Centre for Astronomy and Astrophysics, Ganeshkhind, Post Bag 4, Pune 411007, India}
\email{kanak@iucaa.in}

\begin{abstract}

Estimating the true background in an astronomical image is fundamental to detecting faint sources.
In a typical low-photon count astronomical image, such as in the far and near-ultraviolet wavelength range, conventional methods relying on the 3-sigma clipping and median or mode estimation often fail to capture the true background level accurately. As a consequence, differentiating true sources from noise peaks remain a challenging task. Additionally, in such images, effectively identifying and excluding faint sources during the background estimation process remains crucial, as undetected faint sources could contaminate the background. This results in overestimating the true background and obscuring the detection of very faint sources.
To tackle this problem, we introduce a geometric approach based on the method of steepest descent to identify local minima in an astronomical image. The proposed algorithm based on the minima statistics effectively reduces the confusion between sources and background in the image; thereby ensuring a better background estimation and enhancing the reliability of faint source detection. Our algorithm performs well compared to conventional methods in estimating the background even in crowded field images. In low-photon count, less crowded images, our algorithm recovers the background within 10\%, while traditional methods drastically underestimate it by a few orders of magnitude. In crowded fields, the conventional methods overestimates the background by $\sim 200\%$ whereas our algorithm recovers the true background within $\sim 14\%$. We provide a simple prescription to create a background map using our algorithm and discuss its application in large astronomical surveys.

\end{abstract}


\section{Introduction} 
\label{sec:intro}

From population studies for drug trials to broadband images in astronomy, noise is an inherent property of observations in any research area. Noise in astronomy arises from the nature of data, which mainly depends upon the activity of counting photons from a specific direction in the sky \citep{ODonoghue1995}. Usually, for all wideband CCD photometry \citep{Tyson1984,Jacoby1990} observations, a CCD detector in a telescope is at the focal point, and each pixel in that CCD array is exposed to light from a particular region of the sky. While it collects light from a particular direction, some of the scattered light from other regions may as well enter the pixel due to the scattering in the telescope’s internal optics design, resulting in a spread of point sources to a shape widely known as point spread function (PSF) of the telescope \citep{1958PSF,Racine1996,2023PSF}. 
The number of photons reaching a pixel in a specific interval of time follows a Poisson distribution \citep{Photon_Statistics_1959} where the mean of this distribution is the total time-integrated flux. This process results in a noise known as shot noise \citep{Schottky_1918}. Noise during any astronomical measurement arises from various sources, not only photon counting. An important source is the dark current resulting from the thermal electron movement in the detector, which can introduce fixed patterns or random noise and follow a Poisson distribution, allowing for effective modeling and subtraction during data processing\citep{Howell2006}. During an observation, calibration frames taken with the shutter off are used to correct for the dark current, ensuring accurate separation of the astronomical signal from dark noise. However, besides the dark noise and read noise from the detector electronics \citep{Tyson1984}, other noises are tedious to characterize.

In a dark night observation, we expect some apparently empty regions (considered to be devoid of any source) around an object of interest. Such empty regions are expected to have a non-zero flux which could arise from faint diffuse light from the interstellar medium,
Galactic emission \citep{2000ISM},  Zodiacal light
\citep{Leinertetal1998,Matsumotoetal2018}, Telluric night sky emissions \citep{2008NightSky}, Scattered light from faraway unresolved sources
\citep{Sandin2014},
or the diffuse extragalactic background light \citep{MadauPozzetti2000,Gardneretal2000, Xuetal2005,Murthy2014,EstSkyLvl2018} or stray light from other regions within the field of view (FOV) or outside the FOV \citep{Slateretal2009}. 
This flux, though not the primary focus of this study, is what we refer to as sky or background flux in an image, and ideally, it is subtracted from the flux beneath the sources of interest. The background flux manifests as fluctuations in pixel values around a mean, caused by shot noise, which we refer to as background noise.
The mean value of the background flux ($\Tilde{B}$) may itself vary throughout the FOV, depending on the observation pointing direction and wavelength of the observation. 
The resolution over which the change in the background can be observed depends on the PSF of the telescope. The detection and measurement of fluxes of fainter and distant sources relies heavily on the accuracy of the background measurement.

With the growing necessity of high-precision photometry of extremely low surface brightness galaxies in the local universe or faint distant sources  \citep{KormendyBahcall1974,ImpeyBothun1997,Kodaetal2015,VanDokkumetal2015, PahwaSaha2018,Zaritskyetal2021,Martinetal2022,Bouwensetal2012,Yueetal2014,Ikedaetal2023,2021SahaDhiwarMalin1}, there has been a lot of efforts to measure the sky background as accurately as possible. Several methods have been devised to model the mean background in astronomical images, each of them having its own pros and cons \citep{AkhlaghiIcikawa2015, Kelvinetal2023, Watkinsetal2024, Missing_Light}. Some earlier detection algorithms assumed the noise to be Gaussian \citep{Gausian_BackG_fitting_1993} and modeled it accordingly to find the mode/median of the background, a trend which many newer background estimation techniques have carried forward. The most widely used programs like SExtractor \citep{SExtractor} and Daophot \citep{Daophot_stetson_1987} use an iterative algorithm that divides the image into sub-patches and assume that most of the pixel values belonging to a sub-patch to be dominated by the source-free sky background. 
An iterative $3\sigma$ clipping is used to determine the mode of the background in each sub-patch, as the mean can be skewed by the sources present. The mode is chosen because, for Gaussian distributions, it aligns with the true mean of background and is usually not affected by outliers in data, offering a more accurate estimation of the true background. This method works well for data with high exposure time, such that the mean background in counts $\lambda=\Tilde{B}\times t_{exp}>>1$ per pixel, where $\Tilde{B}$ being the mean background in counts per pixel per unit time, and $t_{exp}$ is the exposure time. In such a case, a Poisson distribution can be approximated with a Gaussian. 

While in the Gaussian regime, it is possible to use an iterative approach, compared to the automated background routines of SExtractor, in that sources are detected first, and then a statistically significant number of boxes can be placed randomly or semi-randomly at locations away from the segmented objects to estimate fluxes in them. From the distribution of these fluxes, one can estimate a mean/median background and noise \citep{Rafelskietal2015, Chayan_GoodsN, Sahaetal2024, Suraj_Dhiwar_LYc}. An issue in this method that remains intractable is the source contamination. This arises due to the faint sources that remained undetected by the chosen SExtractor parameter or boxes that had partial overlap with the sources in the field. 
A way to address this issue would be to subtract an estimated flux contribution of these faint undetected sources from the unsegmented pixels and then estimate the background for detected sources \citep{EstSkyLvl2018}.
In fact, the assumption of an empty region with background dominating over the number of sources is important for the 3-sigma clipping method to sample background pixels. In a crowded field, more such pixels will have contributions from the neighboring sources, which would shift the mode of the background pixel distribution. This effect, known as Confusion limiting \citep{2001Petri_confusion_noise}, is manifested in enhanced background estimation with increasing source density. Besides, the traditional 3-sigma clipping method does not consider the 2D distribution of pixels where a number of pixels from a very faint diffuse source well below the $3\sigma$ limit will be counted as a background sample.    

On the other hand, as $\lambda$ decreases below 10, we are in the regime of medium photon count statistics. For data where $\lambda\approx 4-10$, the Poisson distribution is noticeably right-skewed, making the mode and median less reliable due to the discrete nature of photon counts. 
Given the discrete nature of the Poisson distribution, the median or mode of the distribution always falls at an integer value. For non-integer values of $\lambda$, this integer spacing (representing a minimum count of 1 photon) introduces an uncertainty of approximately $25\%$ to $10\%$ within the range of $\lambda = 4$ to $10$, when using the median or mode as the background measurement instead of the mean.
Due to this skewness and photon counting error, the choice of a mode or median of the background($\Tilde{B}$) distribution would almost always be inaccurate. With $\lambda\approx 1 - 3$, the Gaussian noise approximation would be completely flawed. 
And for $\lambda\approx <0.5$ (low photon count regime), the mode or median of the data would be $0$, leading to $100\%$ error from the mean background flux. Moreover, the ratio of background noise($\sqrt{\lambda}/t_{exp}$) to $\Tilde{B}$ increases as $\lambda$ decreases, making it harder to detect fainter sources. Whether in a crowded field or low-photon counting image, extremely low surface brightness objects such as ultra-diffuse galaxies \citep{VanDokkumetal2015, Zaritskyetal2021} or distant, faint sources might be flagged as spurious sources if the background is overestimated. 

In this paper, we address these issues by devising a new method to sample background pixels by targeting minima in a given 2D image - both in the low photon counting images and in the case of confusion limiting due to high source density. The rest of the paper is organized as follows: section~\ref{sec:minima} describes the properties of minima in an M-dimensional random field drawn from a probability distribution $P(x|\bar{\alpha})$, where $\bar{\alpha}$ are the fixed parameters of the distribution. Section ~\ref{section:method1} deals with directly applying statistics derived in section ~\ref{sec:minima} on images with sources. Section~\ref{sec:crowded} deals with an alternative approach for background sampling in crowded field images and comparing background estimation, while section~\ref{sec:discuss} presents discussion and conclusions derived from this work.

\section{Minima Statistics}
\label{sec:minima}
The Minima are defined in an M-dimensional array as the points having the lowest value with respect to their nearest neighbors. The simplest way to locate a minimum in an M-dimensional random field is to start at a random location in the array and employ the method of steepest descent algorithm, where we start from any pixel and calculate gradients to the nearest neighbors. We shift to the neighboring pixel where this gradient is steepest and non-positive. This is repeated until we reach a pixel where all the gradients to neighboring pixels are positive.

In this section, we analyze the properties of minima from an M-dimensional array of Random variables following an Independent Identical Distribution (IID) and its Probability distribution (referred to as Minima Distribution throughout this paper), with the goal of relating the Minima distribution to the parent distribution. This is achieved by obtaining an isomorphic relation between the CDF of the Minima distribution and the Parent distribution, and by deriving scaling relations between the moments of the Minima distribution and the moments of the parent distribution of the randomly generated M-dimensional array. These relations will later be used to measure true background in Astronomy images.



\subsection{Minima distribution and its parent probability distribution}

All Probability distributions $P(x)$ have Unique Cumulative distribution functions(CDF) $\Phi(x)$ such that $\Phi(x) \equiv P(X< x)$, where the random variable $X$ takes values less than or equal to $x$. And the PDF $P(x)$ is related to the CDF $\Phi(x)$ as $P(x)=\frac{d\Phi}{dx}$

Consider an array of M dimensions whose entries are random numbers, which are realizations of the probability distribution function $P(x)$. In the M dimensional array, we can choose the number of nearest neighbors to be `$n$', and it depends on what positions we are considering for the nearest neighbors with respect to the position of interest. The probability ($P_{min}$) that an entry in the array has a value (say, $x$) and is the local minimum w.r.t. its nearest neighbors can be written as:

\begin{equation}
    P_{min}(x)= N\times P(x)\times (1-\Phi(x))^n,
    \label{eqn:minimapdf}
\end{equation}

\noindent where $P(x)$ is the parent distribution governing the fact that $x$ occurs, and the $\bigl( 1-\Phi(x)\bigr) ^n$ represents the constraint that the $n$ nearest neighbours have higher value than $x$. The normalization constant $N$ is $1+n$ for all continuous probability distribution functions and can be derived as follows:

\begin{align}
    N&= \frac{1}{\int_{x_{min}}^{x_{max}} P(x)\times (1-\Phi(x))^n dx}\nonumber\\
    &= \frac{1}{\int_0^1 d{\Phi}(1-\Phi(x))^n}\nonumber \\
    &= 1+n
\end{align}

\begin{figure*}[ht]
        \centering \includegraphics[width=1\textwidth]{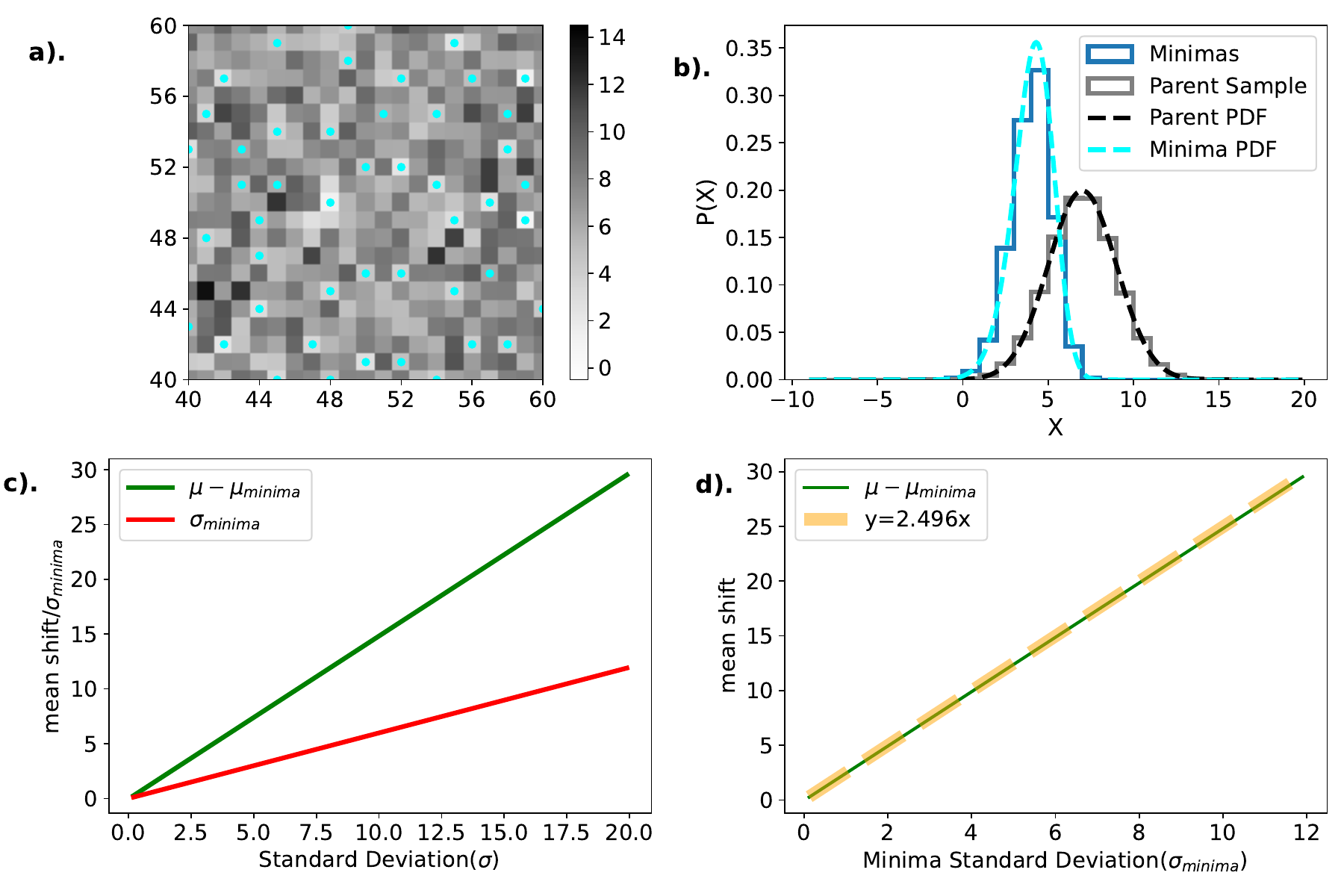}        \caption{\label{Gaussian_ms_s}\textbf{(a)} shows a section of $100\times100$ sample of the generated $2$d Gaussian random field with $\mu=7$ and $\sigma=2$ with the detected minima pixels marked with cyan dots. \textbf{(b)} shows the histogram of all pixels(Parent Sample) and another histogram of detected minima pixels(Minima Sample), plotted together with the theoretical Gaussian distribution(N$[7,\sigma=2]$), and the numerically calculated minima distribution(Minima PDF) calculated from equation \ref{eqn:minimapdf2d}. \textbf{(c)}relation of numerically calculated mean shift($\mu-\mu_{min}$) and $\sigma_{min}$ with $\sigma$ of parent distribution for Gaussian distribution plotted with their respective best fit linear relations. \textbf{(d)} Relations between mean shift with the detected $\sigma_{min}$ and the fit as presented in equation\ref{mu_recoverGaussian}.}
\end{figure*}

\noindent In other words, it is possible to derive an analytic expression for a minima distribution given a parent probability distribution $\bigl(P(x)\bigr)$ and its cumulative distribution function $\bigl(\Phi(x)\bigr)$. A simple test of this is shown in figure \ref{Gaussian_ms_s}, where we generate a 2D random field of 100$\times$100 pixels for a known Gaussian distribution and then detect the minima pixels with respect to their 8 nearest neighbor pixels and over-plotted with the gaussian distribution and its Minima distribution derived from \ref{eqn:minimapdf}. Following this, we show how to recover a parent distribution from its minima distribution.

\noindent We have the relations of PDF and CDF given earlier by $P(x)=d\Phi / dx$ for both minima and the parent distribution. From equation \ref{eqn:minimapdf} we can follow
\begin{align}
    &\quad\quad\quad d\Phi_{min}(x) &=& N\times d\Phi(x)\times (1-\Phi(x))^n\nonumber\\
    &\implies\quad \int_0^{\Phi_{min}}d\Phi_{min}' &=&N\int_0^{\Phi}d\Phi'(1-\Phi')^n\nonumber\\
    &\implies\quad\quad 1-\Phi_{min} &=& (1-\Phi)^{n+1}\nonumber\\
    &\implies \quad\quad\quad\quad\Phi &=& 1-(1-\Phi_{min})^{1/n+1}
\end{align}

Hence, we find that the minima CDF ($\Phi_{min}$) and CDF of parent probability distribution($\Phi$) are isomorphic, and one can be derived from the other in the case of continuous probability distributions.

For the flat background subtraction from an image, the mean of the background pixel distribution is taken. As expected, though, the mean of the minima sample is shifted leftwards from the parent distribution's mean by a constant factor for a particular distribution. This shift ($\mu_{min}-\mu$) is proportional to the standard deviation of the parent distribution. 
\begin{equation}
    \mu-\mu_{min}=C\sigma
    \label{mushift}
\end{equation}

\noindent This constant C is different for different types of distribution (i.e., different for Gaussian, Uniform, gamma, etc). The standard deviation of the minima distribution ($\sigma_{min}$) is also directly proportional to the standard deviation of the parent distribution.

\begin{equation}
    \sigma_{min}=D\sigma
    \label{sigshift}
\end{equation}

Thus the shift in the mean of minima from the mean of parent distribution can be written as
\begin{equation}
    \mu-\mu_{min}=K\sigma_{min}
\end{equation}
where $K=C/D$. and thus, we can recover the parent distribution's mean from the minima distribution mean by 

\begin{equation}
    \mu=\mu_{min}+K\sigma_{min}
    \label{mu_recover}
\end{equation}

\subsection{Minima moments in 2D random fields}
The Probability distribution function of the minima in a 2D array of independent and identically distributed random variables, where the number of nearest neighbors is 8, is given by

\begin{equation}
    P_{min}(x)=9P(x)(1-\Phi(x))^8
    \label{eqn:minimapdf2d}
\end{equation}

An example test of the above relation (equation \ref{eqn:minimapdf2d}) is shown in figure \ref{Gaussian_ms_s}(a) and \ref{Gaussian_ms_s}(b) for a given Gaussian field.
Further, the 1st and 2nd moments of the minima distribution are given by
\begin{align}
    \mu_{min}&= \int_{-\infty}^{\infty} x 9P(x)(1-\Phi(x))^8 dx\label{mumin1}\\
    \sigma^2_{min}&= \int_{-\infty}^{\infty} x^2 9P(x)(1-\Phi(x))^8 dx - \mu_{min}^2
\end{align}

We can write $Pdx=d\Phi$ and since $\Phi(x)$ is a continuous monotonously increasing function of $x$ in its domain, it can be inverted, and $x$ can be written as a function of $\Phi$. so the above equations become.
\begin{align}
    \mu_{min}&= 9\int_{0}^{1} x(\Phi) (1-\Phi)^8 d\Phi\\
    \sigma_{min}^2&= 9\int_{0}^{1} x(\Phi)^2 (1-\Phi)^8 d\Phi - \mu_{min}^2
\end{align}

\medskip \noindent \textbf{For Gaussain Distribution:}

In case of Gaussian PDF with mean $\mu$ and standard deviation $\sigma$,

\begin{equation}
    2\Phi -1 = Erf\left(\frac{x-\mu}{\sqrt{2}\sigma}\right)
\end{equation}
Inverting for $x$, we get the mean shift and minima standard deviation  given by

\begin{align}
    \mu-\mu_{min} =\biggl(-9\sqrt{2}\int_{0}^{1} Erf^{-1}\left( 2\Phi -1\right)(\Phi)\nonumber\\ \times (1-\Phi)^8 d\Phi\biggr)\sigma
    \label{mushiftgaussian}
\end{align}

\begin{align}
    \sigma_{min}=\biggl(-18\int_{0}^{1} \left(Erf^{-1}( 2\Phi -1)\right)^2 (1-\Phi)^8 d\Phi\nonumber\\
    -(\mu-\mu_{min})^2 \biggr)^{1/2} \sigma
    \label{sigshiftgaussian}
\end{align}

Computing $C_{gaussian}$ and $D_{gaussian}$ from equations \ref{mushift}, \ref{sigshift}, \ref{mushiftgaussian} and \ref{sigshiftgaussian} gives $C_{gaussian} = 1.48501$ and $D_{gaussian}=0.59779$ which are close to the Values obtained by straight line fitting in figure \ref{Gaussian_ms_s}, which are $C_{gaussian}=1.486 \pm 0.006 $ and $D_{gaussian}=0.596 \pm 0.005$ .


So for estimating the parent distribution's mean and sigma from the minima from eq \ref{mu_recover}, we estimate $K=C/D$ as $\approx 2.496$, so we can write the mean recovery equation for a Gaussian field as 
\begin{equation}
    \mu=\mu_{min}+2.496\sigma_{min}
    \label{mu_recoverGaussian}
\end{equation}

\begin{figure*}[ht]
    \centering 
    \includegraphics[width=1.0\textwidth]{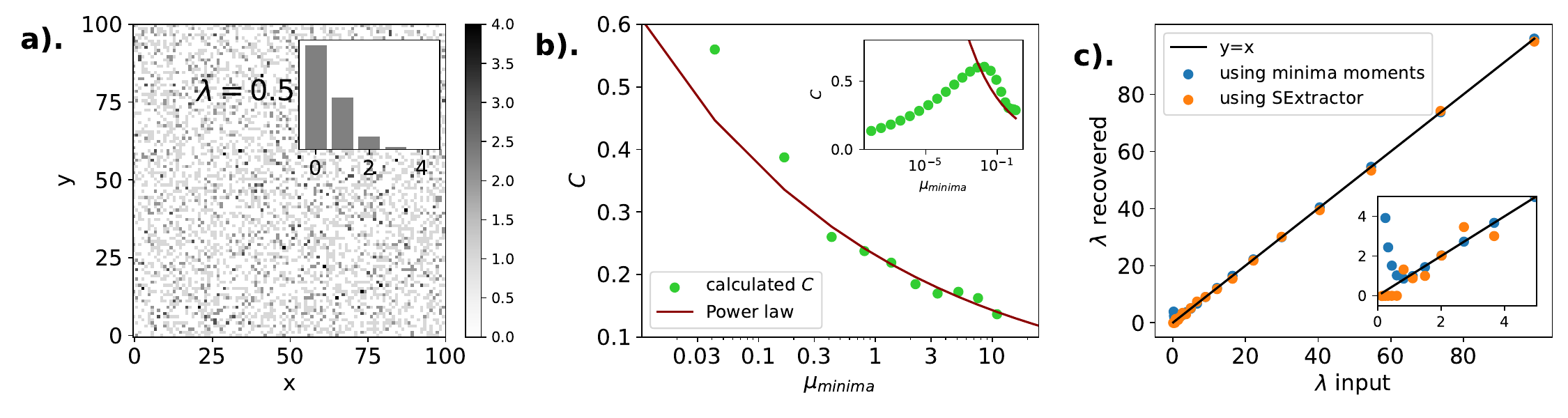}
    \caption{\label{fig:correction_f}$(a)$Generated Poisson Random Variable field example, with its pixel histogram as an inset figure, and the parameter $\lambda$ mentioned in the top left corner  $(b)$ Numerically calculated Correction factor for the Poisson minima relation in equation $\ref{lam_recover}$, with the power law approximation which holds for $\mu_{min}\gtrsim 0.3$ $(c)$ Recovered $\lambda$ from generated Poisson random fields using equation $\ref{lam_recover}$ and recovery is compared with the mean background estimations by Source Extractor. }
\end{figure*}

\medskip \noindent \textbf{For Poisson Distribution:}

A Poisson distribution with mean $\lambda$ is given by,
\begin{equation}
    P(x|\lambda)=e^{-\lambda}\frac{\lambda^x}{x!}
\end{equation}
where $x\in \{0\} \bigcup \mathbb{Z}^+$.
While the high mean Poisson distribution is usually approximated with Gaussian, the low mean Poisson distribution varies in its shape and properties from the Gaussian by a huge margin. For $\lambda<<1$, the $\mu_{min}$ recovered from equation \ref{mumin1}, is dominated by $x=1$ term, and we get $\mu_{min}\approx \lambda^{9}$ which goes down to zero quickly as $\lambda$ decreases.

Using the Gaussian constant $K=2.496$ for relating the $\mu_{min}$ and $\sigma_{min}$ to $\lambda$ in the case of Poisson field, we have to account for a correction factor $c$. This correction factor is modeled with an empirical power law (figure \ref{fig:correction_f}) over $\mu_{min}$, which becomes negligible as $\lambda$ increases. The $\lambda$ recovery from $\mu_{min}$ recovered becomes the equation \ref{lam_recover}

\begin{equation}
    \lambda\approx \mu_{min} + 2.496\sigma_{min} + A\mu_{min}^{\gamma}
    \label{lam_recover}
\end{equation}
We calculate $A$ and $\gamma$ numerically and find $A\approx 0.231$ and $\gamma\approx -0.2085$.

\subsection{Tests on simulated Poisson fields}
To test the relation obtained in equation \ref{lam_recover}, we generate Poisson random variable images for simulating Poisson fields (flat images with shot noise); we use the pixel units of counts per pixel throughout this paper.
We generated 300$\times$300 pixel images of Poisson random variables (example in figure \ref{fig:correction_f} a), from $\lambda=0.15$ to $100$, to test equation \ref{lam_recover}'s reliability. 
We sampled all the minimum of the images using the steepest descent method and calculated $\mu_{min}$ by averaging over the samples. The resultant $\lambda$ recovery is shown in figure \ref{fig:correction_f}(c), which is compared with the mean background calculated by SExtractor's background routine. It is clear that both methods do well at the high $\lambda$ limit, but the lambda recovery using equation \ref{lam_recover} is relatively more accurate at the $\lambda\gtrsim 1$ limit. However, we see that for $\lambda<1$, both methods produce wrong estimates, where SExtractor gives values closer to $0$, the minima recovery equation produces overestimation due to the correction factor's power law model which is inaccurate at $\mu_{min}<0.3$ and by extension low $\lambda$ (figure \ref{fig:correction_f} b and c).

In the next section, we see the applicability of these minima recovery equations in the mean background estimation of images with faint point sources.
\begin{figure*}[ht]
        \centering \includegraphics[width=1.0\textwidth]{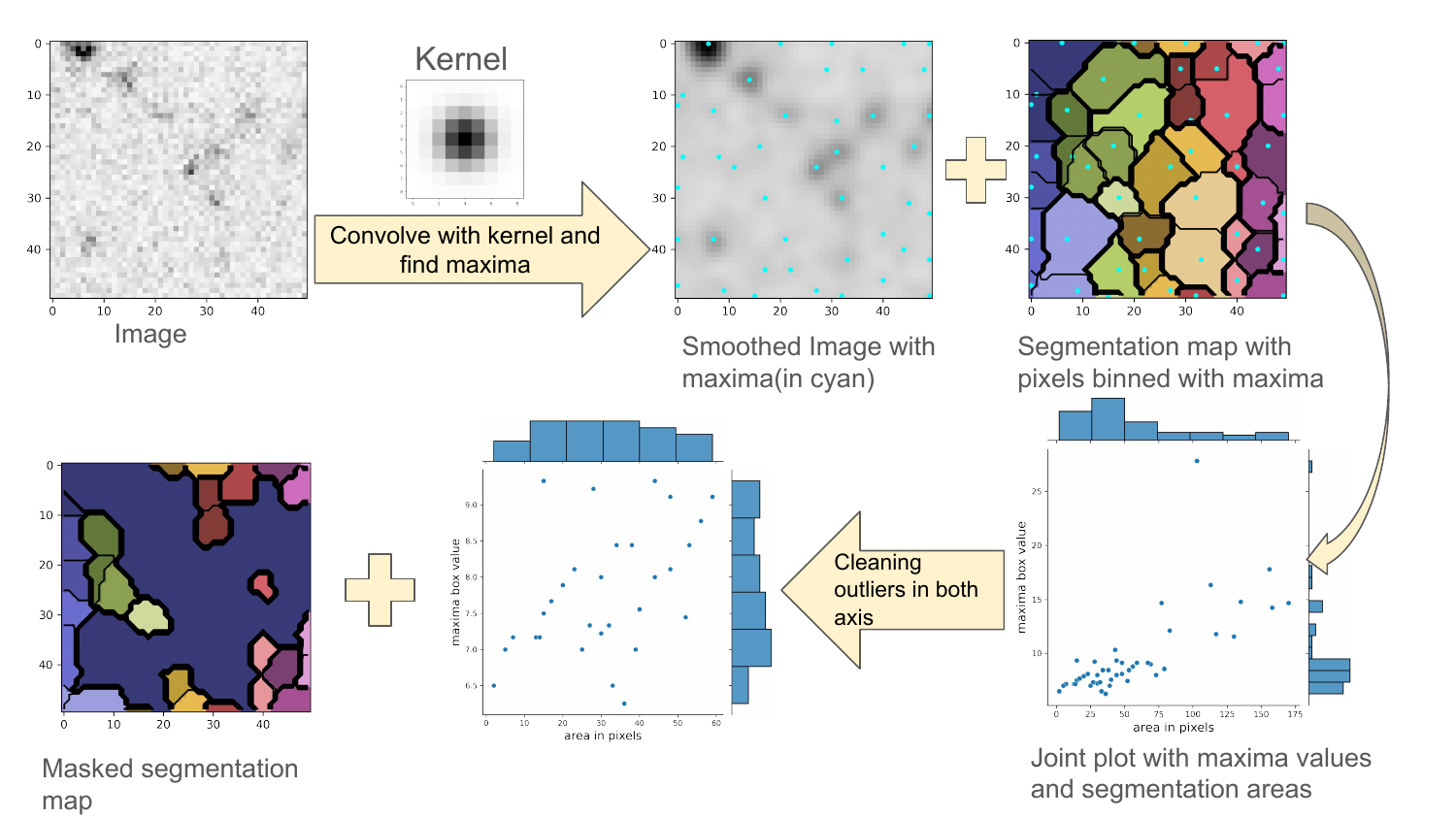}
        \caption{\label{fig:MaskingFlowchart}A simplified flowchart showing the process of maxima-based masking of regions with sources}
\end{figure*}

\section{Application of Minima sampling on images with sources}
\label{section:method1}

The tests in the previous section were done on source-less Poisson random fields.
Sources can be defined as clumps of high values (compared to the background Poissonian noise) in the 2D image. These clumps could also occur spuriously from noise, but the spurious clumping has a very low probability of occurring. For example, A clump of $n$ pixels above 1 sigma from the background noise has a occurring probability of $\sim \bigl(1-\Phi(\mu+\sigma)\bigr)^n$, For a Gaussian variable, it is $\sim 0.16^n$, which goes down in orders of magnitude with $n$.

To simulate Images with sources, we take UVIT-N242w pixel scale and psf as reference \citep{UVIT_Pipeline_Tandon2022}, we injected point sources with Gaussian point spread function(with fwhm$\sim$ 2.87 pixels or 1.2 arcseconds) at random locations in a 300$\times$300 pixel (2.085 $\times$ 2.085 arcminutes$^2$) image, and the sources uniformly having SNR (within psf fwhm) from 1 to 7 (figure \ref{fig:lamrec_image_w_sources}) with respect to background, (for our test case we take background $\lambda=6.3$).We made images with a number of sources ranging from 300 to 3600 (69 to 828 sources/arcminutes$^2$), which look similar to the images shown in figure \ref{fig:lamrec_image_w_sources} a. 

Now, to calculate the Background from these simulated images, we came up with a way to mask sources, otherwise some of the detected minimums may get lodged between intersection of sources, they could also lie on relatively flat sources, resulting in an increase in the recovered mean background value.

\subsection{Source masking and mean background recovery routine}

For source masking, we use the Fellwalker \citep{Fellwalker_berry_2015} algorithm to identify maxima on the simulated image convolved with a smoothing kernel of our choice, which should be nearly around the psf of these fields. In the smoothed image, the maxima were located, and the whole field was segmented, where all the pixels leading to a maximum were assigned to the same maximum, similar to the process in \cite{Fellwalker_berry_2015}, but rather than starting from a threshold, we segment all the pixels. We then calculate the segment area (in pixels) and maximum flux with a $3\times 3$ box at the position of maximum simultaneously. We assume that the bins containing sources have higher area and maximum flux values, thus distinguishing them from those randomly arising from background pixels. We perform masking of the source bins by 3 sigma clipping on both maxima flux and bin area. The remaining bins are then used to calculate the background. A simplified flowchart showing the whole process is presented in figure \ref{fig:MaskingFlowchart}.

The automated routine applied for the tests on images with sources is as follows
\begin{itemize}
    \item Apply a simple Gaussian smoothing on the image not exceeding the psf to simplify source detection.
    \item Detect maxima in the smoothed image, number them, and mark the pixels leading to a maximum by the steepest descent method by the same number, creating a segmentation map of the smoothed image based on maxima.
    \item Identify the maxima segmentations containing possible sources and mask them (fig. \ref{fig:MaskingFlowchart}).
    \item Identify minima on the unmasked regions of the original unsmoothed image and recover lambda from equation \ref{lam_recover}.
\end{itemize}
\subsection{Test Results and limitations}

\begin{figure*}[ht]
\hspace{-1cm}
        \centering \includegraphics[width=1.0\textwidth]{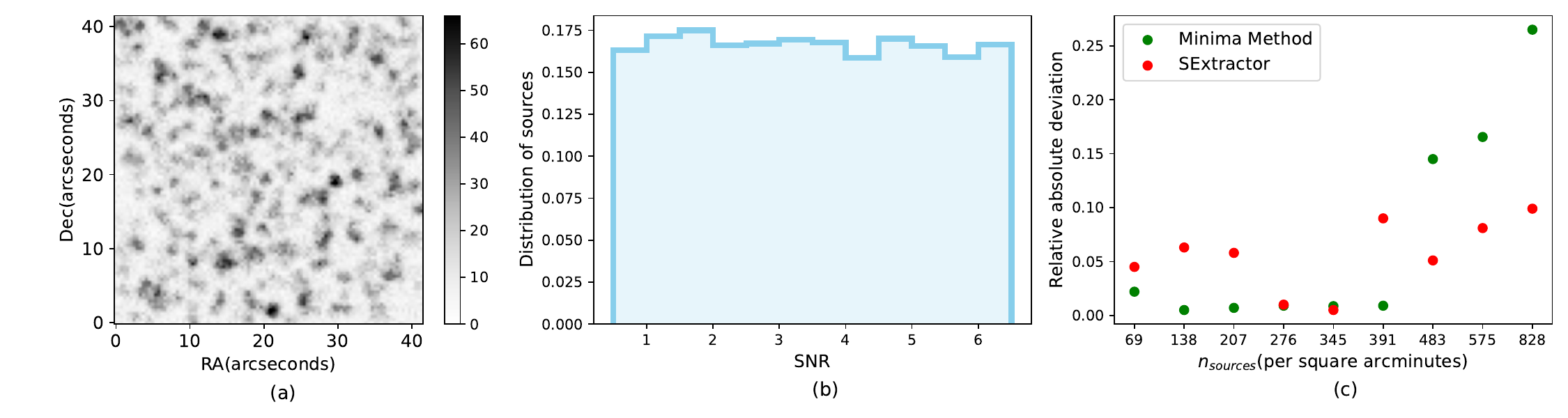}
        \caption{\label{fig:lamrec_image_w_sources} \textbf{(a)} A sample Poissonian image with point sources injected as per the method described in section \ref{section:method1}. \textbf{(b)} SNR distribution of the injected sources of one of the generated images. While the mean background and standard deviation of the images change in tests, the flux of the sources is chosen such that their SNR distribution remains the same. \textbf{(c)} Relative absolute deviation of recovered background from crowded fields from the source masked images for the tests in section \ref{section:method1} using the equation\ref{lam_recover}, in comparison to SExtractor background}
\end{figure*}

The test results are presented as a relative deviation plot in figure \ref{fig:lamrec_image_w_sources}, where the y-axis represents the relative absolute deviation (equation \ref{eqn:deviation}) of recovered background in the generated image from the input background.

\begin{equation}
     \text{Relative absolute deviation}= \frac{|(\lambda_{input}-\lambda_{recovered}|}{\lambda_{input}}
    \label{eqn:deviation}
\end{equation}

From figure \ref{fig:lamrec_image_w_sources}, we find that the background recovery from this minima sampling routine works well within a relative absolute deviation of $\lesssim 0.2$ till a source coverage of 391 point sources per square arcminutes (or 1700 point sources in a 300$\times$300 pixel canvas with a pixel scale of 0.417 arcseconds). A simple calculation for this amount of sources and assuming every point source's average span ($>90\%$ flux) is contained within 5$\times$5 pixels (as psf FWHM is 2.8 pixels), we find that approximately $\sim 50\%$ of the image is covered by source pixels. Increasing sources beyond this limit causes the minima routine to not identify sources accurately for masking, and hence, the recovered background shows a high deviation from the input background value.

We have shown the comparison of background deviation from our minima sampling routine with the source extractor's background routine in figure \ref{fig:lamrec_image_w_sources}(c). We note that after reaching $\gtrsim 50\%$ source coverage, the recovered background by our method diverges much more than the source extractor. 

The above method, as discussed before, has the following notable limitations, which are as follows.
\begin{itemize}
    \item It cannot reliably probe the background of images with $\lambda=\Tilde{B}\times t_{exp}$ less than 1, as most notable minima will be 0's, which are unscalable.
    \item It requires a source masking step before it can start sampling background minima pixels, which is accurate if sources are very bright in comparison to the background, but faint point-like sources may not be identified and masked accurately.
    \item Also, in crowded images where the source coverage of the image is very high ($\gtrsim$ 50\% area of the image), the source masking technique fails to determine the sources correctly, which results in an overestimation of the background.
    \item All tests were performed on idealized synthetic images with generated flat noisy background following an IID, while real images often contain correlated noise due to artifacts (bad pixel columns, cosmic ray hits, saturation bleeds, chip gaps, satellite trails, etc.) and interpolation techniques,  complicating background estimation as they won't follow the relations derived in the previous section for independent and identically distributed noise. 
\end{itemize}

While theoretically, by the minima statistics, we would be able to probe the background of an astronomy image, it seems if the field is crowded or the background $\lambda\lesssim 1$, the aforementioned method fails in accurately determining the background value. Also, to bypass the source masking step before we calculate the background, we devise another method in the next section, which prioritizes finding relatively empty regions with the help of minima in the smoothed images for background calculation.

\section{background estimation for crowded and low $\lambda$ fields}
\label{sec:crowded}

Here we present a second approach to address the limitations we encountered in section \ref{section:method1}. We explore the intuition that minima in a smoothed observation image are least affected by sources and, hence, are good candidates for locating regions to sample background. The base noisy observation image $\bigl( I(x,y)\bigr)$ is convolved with a known normalized kernel $G(x,y)$,  resulting in a smoothed image ($I_{smooth}$).

\begin{equation}
    \begin{centering}
        I_{smooth}=G*I
    \end{centering}
\end{equation}

 We sample the region around the local minimum locations (from $I_{smooth}$) in $I$ with a box of size $b$. We find in the case of Gaussian random variables in the base image, the pixels sampled at and around a minimum hold a similar relation as equation \ref{mu_recover}, only the coefficient $K$ (here we'll call it $F$) is now dependent on the smoothing kernel ($G$), and the pixel sampling box length ($b$) at the minima positions (equation \ref{smoth_lambdarec}).

\begin{equation}
    \mu - \mu_{min} = F(G,b)\sigma_{min}
    \label{smoth_lambdarec}
\end{equation}

If $b$ is large enough ($\gtrsim$ FWHM of $G$), we find $\sigma_{min}\approx \sigma$. More details on the $F(G,b)$ dependence on its parameters can be found in section \ref{section:Appendix_A} of Appendix. Smoothing the image is an important step, as it connects the disjointed source pixels in case of high noise. We assume the minima in the smoothed image to be good starting points for finding relatively empty (source-free) regions. But just using one smoothing kernel for convolving the image presents its own limitations.
\begin{itemize}
    \item In the case of a narrow smoothing kernel, some of the local minima on $I_{smooth}$ can lie in between bright sources; hence, sampling in the minima neighborhood may cause us to pick some pixels from those sources.
    \item Using a large/wide smoothing kernel allows most minima to be located far away from bright sources, but it may cause fainter sources in empty regions to merge with local minima, which can again cause us to pick pixels from possible sources. Also, the number of minima locations on a $I_{smooth}$ with large $G$ are significantly less, which can lead to low sampling.
\end{itemize}

While choosing the sampling box length $b$, we cannot increase it arbitrarily since a large sampling box may intrude into regions occupied by sources. We develop an iterative method that uses the minima in images smoothed with large $G$ to find regions away from obvious bright sources, then iteratively find minima near these regions in the images smoothed with smaller $G$, which would indicate the source-free regions. The details of this recipe are presented in the next subsection.

\subsection{Recipe to locate empty regions in crowded images}

\begin{figure}
    \centering 
        \includegraphics[width=1.0\columnwidth]{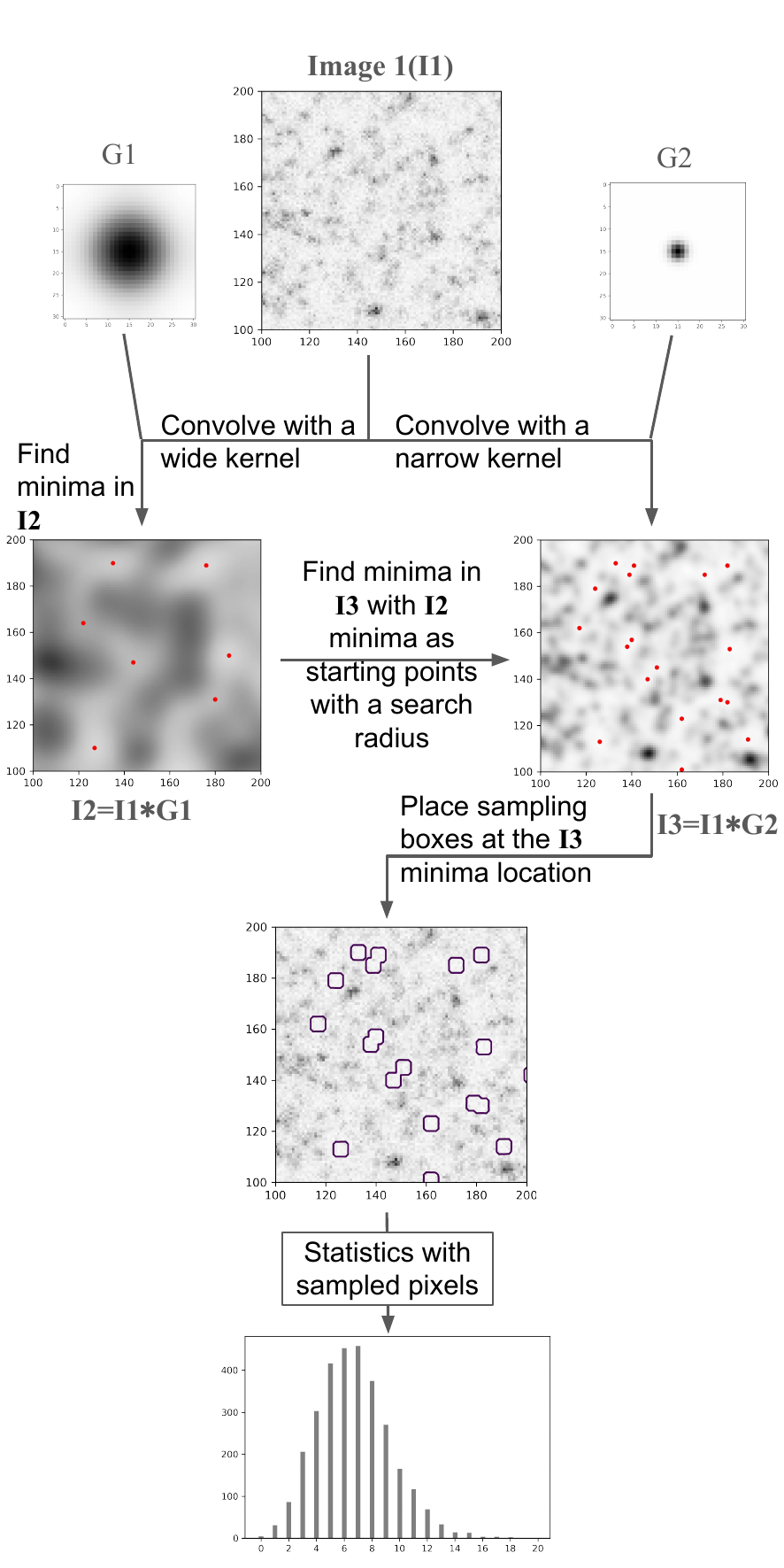}
        \caption{\label{fig:Minima2Flowchart}A simplified flowchart showing the process of minima-based background sampling, with the detected minima in the intermediate stage smoothed images (I2 and I3 ) shown by red points and the sampling boxes marked by purple in the second-to-last step. NOTE: The transition from I2 to I3 happens over a few iterative stages with smoothing Kernels having widths in between $G_1$ to $G_2$. }
\end{figure}

We use Gaussian kernel-smoothed images to search for minima. These local minima are expected to be less influenced by local sources. However, such minima can occur at the junction of bright sources, and sampling at these minima positions can influence the background estimation. Also, near faint objects, the size of the Gaussian kernel largely dictates the minima positions, so we have set up a recipe to reliably locate minima in empty regions away from bright sources for background sample collection. The employed recipe is listed below and shown in a simplified manner in figure \ref{fig:Minima2Flowchart}.

\begin{enumerate}
    \item Convolving the image with a wide FWHM (usually 5-10 times the psf FWHM) Gaussian kernel and locating the minima via steepest descent in the resultant image. This step should take care of the sampling location being far from bright sources.
    \item Now, the original image is convolved with a Gaussian kernel of slightly narrower FWHM than the previous step, and the pixels around the previous minima are taken as starting positions for the search for new minima. 
    \item Repeat step (2) multiple times till we reach the smoothing kernel size of (1-2 times psf). Also, note that we should use all the pixels nearby ($\gtrsim$ 2$\times$PSF) previous minima to locate other minima for getting more samples. This iteration step helps in increasing the number of minimums obtained for background analysis.
    \item After obtaining the final list of minima coordinates, we put a box of specific length (about the size of last smoothing kernel's FWHM), centered on the minima position on the original unsmoothed image to sample the background pixels around it while also avoiding double counting of pixels.
    
\end{enumerate}

Here, we have set the goal of the method to keep source influence to a minimum. We test the method on a series of generated observations (primarily on source-dense/confusion-limited images and low photon count images). 

\subsection{Tests on random fields without sources}
To test this new minima sampling routine, we ran tests on source-free empty images with a mean background 0 and Gaussian noise and further tests with shot noise for a given $\lambda$. For the tests, all generated images have pixel units of counts pixel$^{-1}$. While the algorithm can be readily applied to images with units of counts pixel$^{-1}$ second$^{-1}$ in case of Gaussian approximation of noise, but for low photon count images, the distribution of noise is not Gaussian and requires some correction factors over the Gaussian relations which need adjusting by taking the exposure time into consideration.

\medskip
\noindent \textbf{In Gaussian random fields:}
\\
We ran the above routine on a set of generated $300\times 300$ pixel images, whose pixels were Gaussian random variables following $N\left(0,\sigma^2\right)$, where $\sigma$ was varied from 2 to 50, We fix the starting smoothing kernel fwhm to \textbf{15.3} pixels, final smoothing kernel to 3.5 pixels, with 6 smoothing steps where the routine gradually decreases the smoothing kernel radius from starting to final sigma, and we fixed the sampling box size ($b$) to 7 pixels. We find the standard deviation of sampled pixels from the above routine $\sigma_{min}$ to be equal to the $\sigma$ of the Normal distribution used. But the mean of sampled pixels followed an empirical relation shown in equation \ref{smoth_murec_test}.

\begin{equation}
    \mu - \mu_{min} = (0.213\pm 0.002)\sigma
    \label{smoth_murec_test}
\end{equation}

We must keep in mind that this coefficient 0.213 is fixed for the set of parameters that we used for the test, but if the parameters like initial and final smoothing kernels ($G_{1},G_{2}$) and box size $b$ is changed, the coefficient differs with a stronger inverse relation to $b$, and a weaker inverse relation to $G_{2}$ when $G_{2\ FWHM} \gtrsim b$ and weak positive relation for $G_{2\ FWHM} \lesssim b$. A further discussion on this coefficient's dependencies on its parameters is presented in Appendix \ref{section:Appendix_A}.

\begin{figure}
    \centering
    \includegraphics[width=1.0\columnwidth]{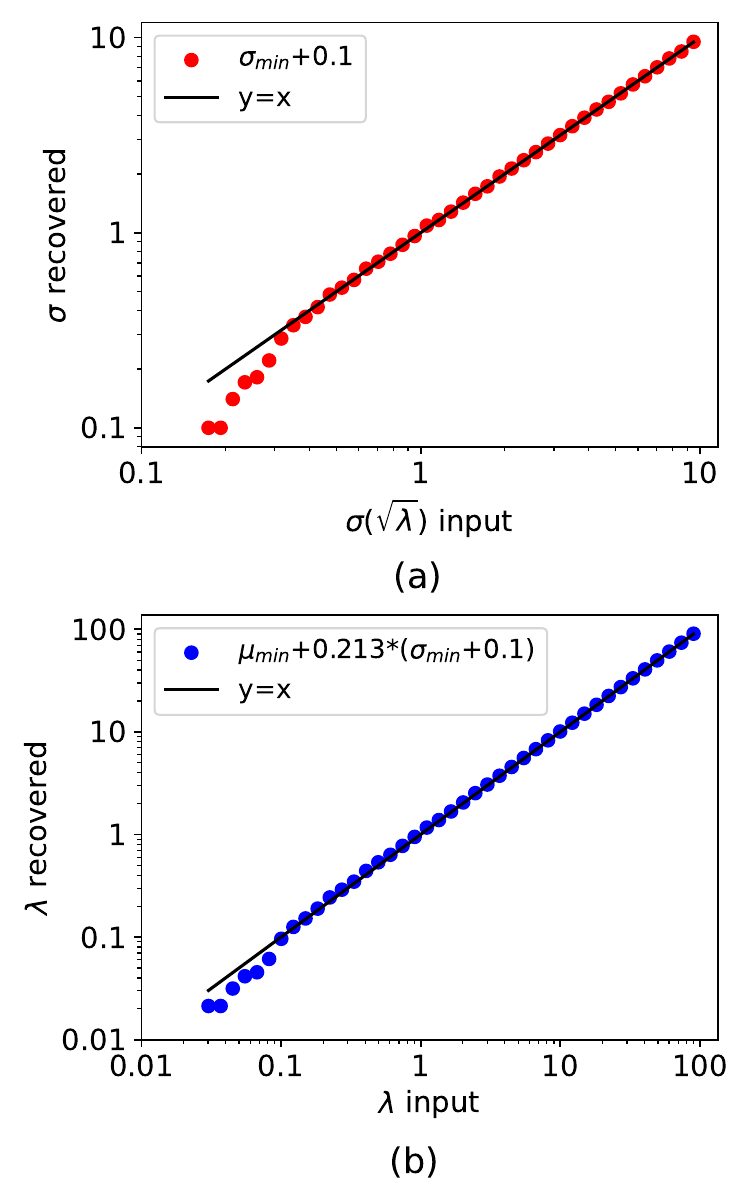}

    \caption{\textbf{(a)} We show the least square fitted linear relation between $\sigma_{min}$, calculated from sampled pixels from Poisson images and $\sqrt{\lambda}$. \textbf{(b)} The fitted relation between $\mu_{min}$ and input $\lambda$ in the Poisson images.}\label{fig:graphs}
\end{figure}

\medskip
\noindent \textbf{In Poisson random fields:}
\\
We again ran the routine on a set of generated Poisson images with $\lambda$ increasing from $\sim$ 0.03 to 90. Using the same parameters for the routine, we get the empirical relations between $\sigma_{min}$ and $\sigma=\sqrt{\lambda}$ to be (figure \ref{fig:graphs} a)

\begin{equation}
    \sqrt{\lambda} = \sigma_{min}+0.1
    \label{smoth_murec_test_pois1}
\end{equation}

and the empirical relation between $\mu_{min}$ and $\lambda$ holds similarly with equation \ref{smoth_murec_test} (figure \ref{fig:graphs} b)

\begin{equation}
    \lambda_{recovery} = \mu_{min}+0.213(\sigma_{min}+0.1)
    \label{smoth_murec_test_pois2}
\end{equation}

From the figure \ref{fig:graphs}, we can note that here we recover $\lambda $ of the order of $\sim 0.1$ with good accuracy, which was a limitation in the minima method in the previous section. For images in units of counts second$^{-1}$ pixel$^{-1}$ (cps per pixel), the only change required in $\sigma$ correction presented in equation \ref{smoth_murec_test_pois1} is $\sqrt{\lambda}/t_{exp} =\sigma_{min} + 0.1/t_{exp}$.

\subsection{Tests on simulated Poisson fields with sources}

\begin{figure}
    \centering
    \includegraphics[width=\columnwidth]{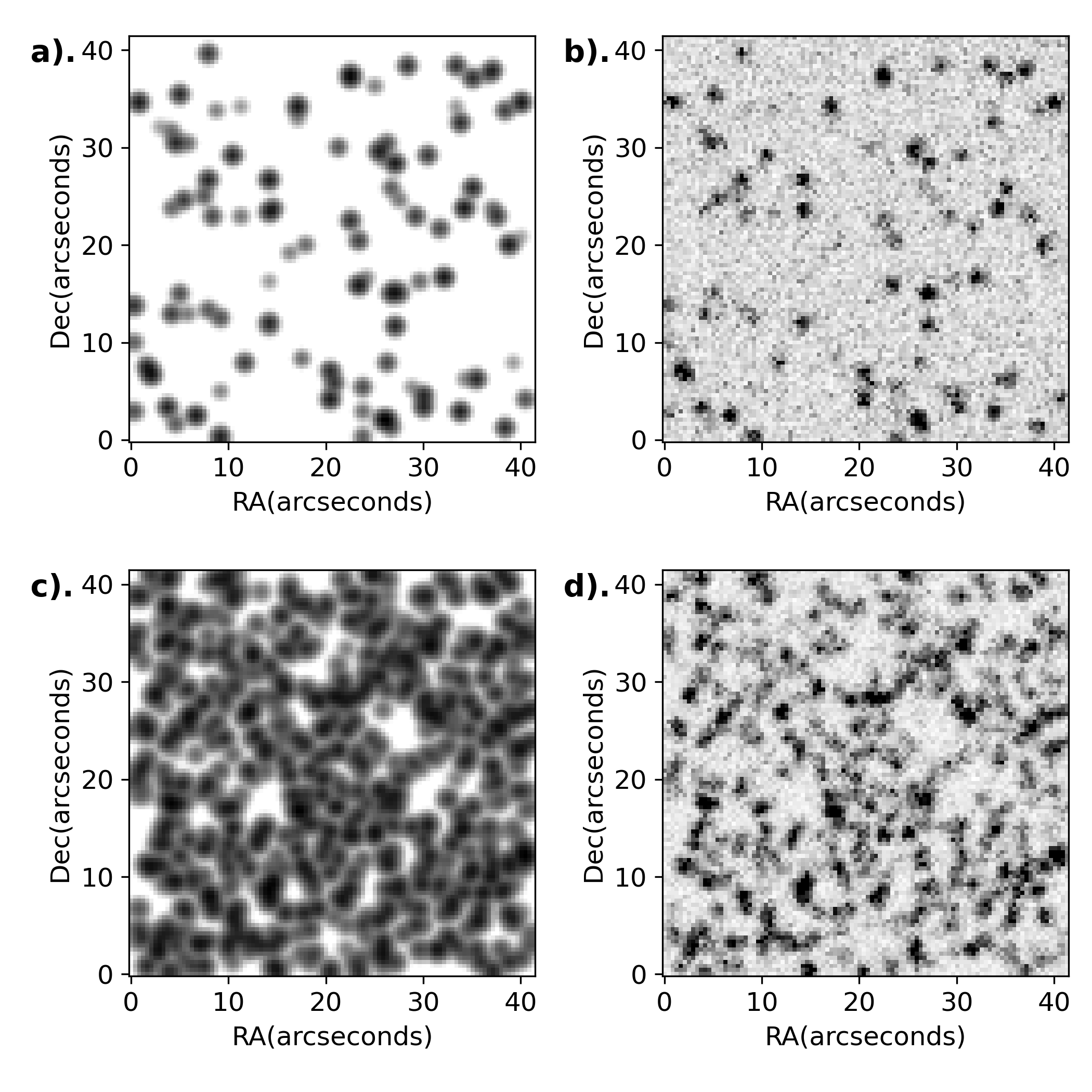}

    \caption{Here we show a sample of generated Image with sources \textbf{(a)} a $100\times 100$ pixel($\sim41\times41$ arcsecond$^2$) section of $300\times300$ pixel canvas where 500 Point Sources are uniformly distributed with a finite Gaussian psf \textbf{(b)} shows the same sources in (a) lying on a noisy background. \textbf{(c)} a $100\times 100$ pixel($\sim41\times41$ arcsecond$^2$) section of $300\times300$ pixel canvas where 4500 Point Sources are uniformly distributed with a finite Gaussian psf and \textbf{(d)}shows the same sources in (c) lying on a noisy background. We can see that the source field in (a) is less crowded, and the source field in (d) is highly crowded.} 
    \label{fig:source_pic_sample}
\end{figure}

Now to test the minima routine on Images with injected sources, we again generate test images with point sources in a similar way as we did in section \ref{section:method1}. But this time, the test was run on images containing 500, 2000 and 4000 thousand sources, and using a Gaussian psf with 2.8-pixel fwhm for emulating the crowding effect as shown in figure \ref{fig:source_pic_sample}. The goal of this routine is to conservatively select regions that are source-free or the least affected by source flux in a local neighborhood. To test this, we run the Minima routine on these generated images with background mean $\lambda$ varying from $\sim$ 0.1 to 100 counts, and for each generated image, again, the generated source fluxes ranged uniformly between 0.7 to 7 SNR (within the psf fwhm) which are calculated with respect to the input background.

\begin{figure*}[ht]
    \centering
    \includegraphics[width=0.9\textwidth]{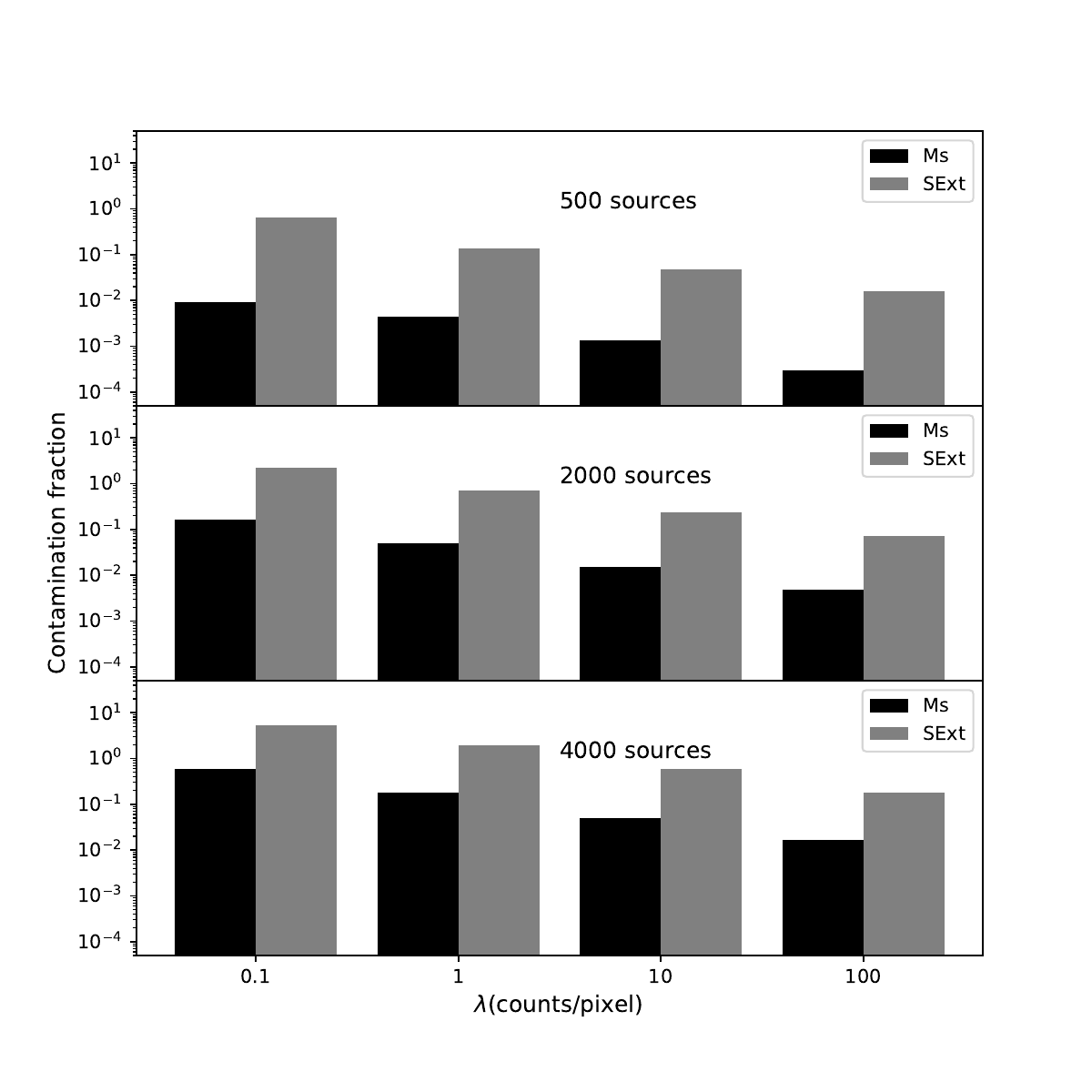}

    \caption{Checking the contamination fraction in our generated images. from top to bottom, we increase the number of sources going from relatively uncrowded to very crowded. And from left to right, we increase the mean background($\lambda$), which goes from the low count Poisson to the Gaussian limit of Poisson distribution} 
    \label{fig:sourc_cont_bar}
\end{figure*}

We define a Source Contamination term ($S_c$) as the average flux contribution of photon counts to the background sample by sources in the field. When a background sample is drawn from the Astronomy image, we expect the present sources to have some photon contribution to the sampled background pixels.
This average source flux contribution ($S_c$) adds another complexity to the background recovery as it must be subtracted from the obtained Background sample before adding the correction factor discussed in the previous section. The background lambda recovery equation in equation \ref{smoth_lambdarec} becoming 

\begin{equation}
    \mu = \mu_{min}+F(G,b)\sigma_{min}-S_c
    \label{smoth_murec_test_Sc}
\end{equation}

Since $S_c$ cannot be estimated independently of the background, we need to minimize it with our background sampling method. We tested it for the generated images with sources. We define the Contamination fraction as 
\begin{equation}
    \text{Contamination fraction}=\frac{S_c}{\lambda}
\end{equation}
Where this Contamination fraction can be considered the fraction, by which the background recovery would be overestimated due to the sources present in this test.

We test this by running the minima sampling routine on the generated noisy point source images and using the same sampling boxes on background-free and noise-free source images.
The mean flux captured by sampling from the source images is divided by the input background value to calculate the contamination fraction. We also calculate the contamination fraction of SExtractor's background routine by sampling at the locations in the source fields, where the pixels of the noisy image are under $\mu_{SEx}+3\sigma_{rms}$ where $\mu_{SEx},\sigma_{rms}$ are mean background and background rms, calculated by SExtractor. 

We show the contamination fraction of the Minima sapling routine and SExtractor in figure \ref{fig:sourc_cont_bar} on the same generated images. We see that in the case of relatively uncrowded images, the source contamination is lower for both SExtractor and Minima Sampling, but in very crowded images, source contamination increases, which is as expected. This trend is seen for all images irrespective of the mean $\lambda$. 
However, the contamination fraction of the Minima sampling method stays below 1 even in the case of very crowded fields, and it stays lower by 1-2 orders of magnitude compared to SExtractor in all cases. This suggests that the Minima sampling method is able to avoid sources better than the $3\sigma$ clipping algorithm of SExtractor.

\begin{figure*}[ht]
\centering
    \includegraphics[width=1\textwidth]{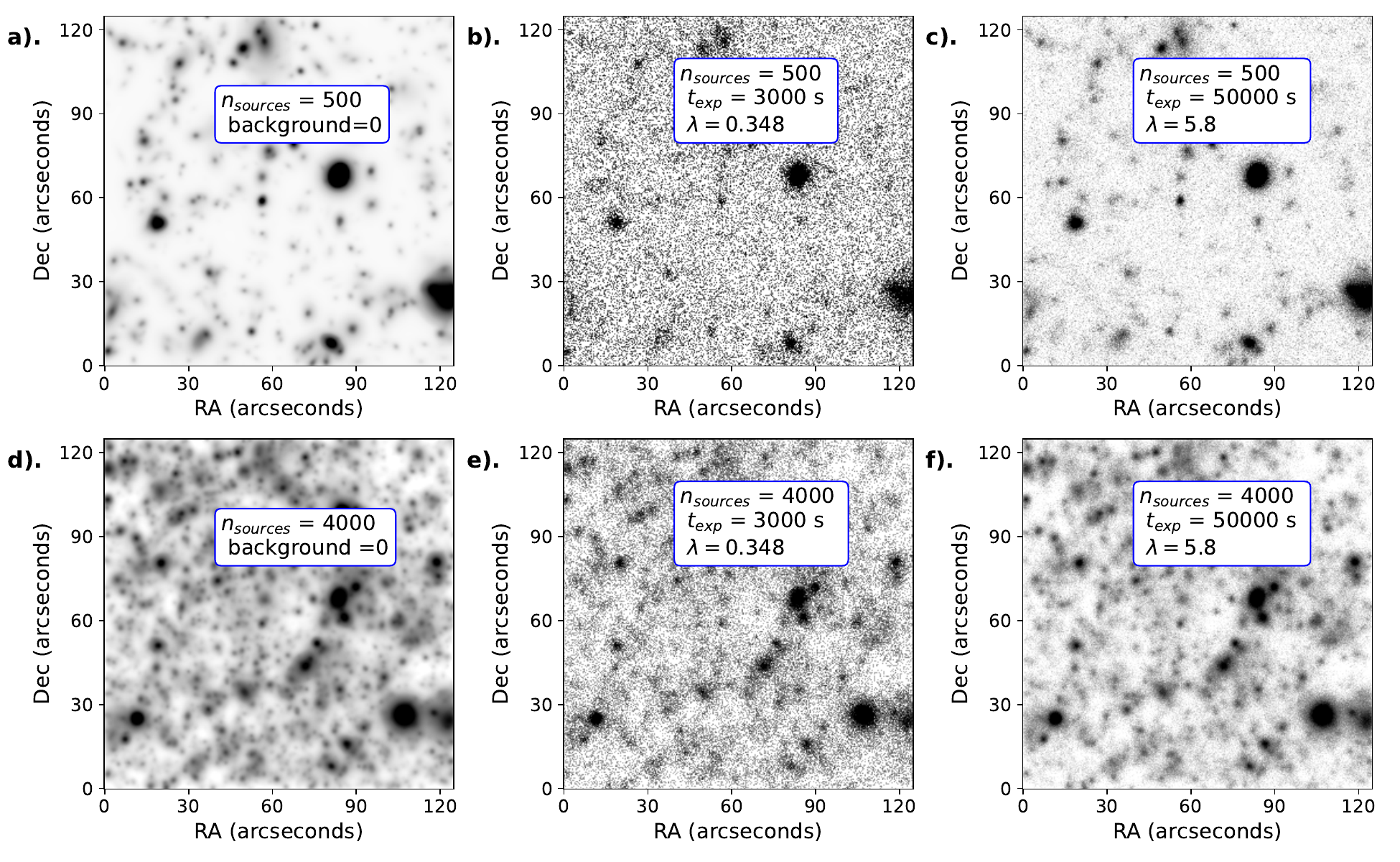}
    \caption{Sample generated UVIT-like images. \textbf{(a)}Background and noise free 2.085$\times$2.085 arcminute$^2$ (300$\times$300 pixels$^2$) Field with $\sim$ 500 sources ($\sim$ 125 per square arcminute ). \textbf{(b)} and \textbf{(c)} shows the same field as (a) with exposure times of 3000 and 50,000 seconds, respectively, at a mean background flux of $1.16 \times 10^{-4}$ cps per pixel. \textbf{(d)} Background and noise free field with $\sim$4000 sources $(\sim$500 per square arcminute). \textbf{(e)} and \textbf{(f)} show the same field as (d) with exposure times of 3000 and 50,000 seconds, respectively, at the same background flux.}
    \label{fig:fake_uvit_samp}
\end{figure*}

\subsection{Tests on simulated UVIT-like Images}
To test the Minima Sampling method on images closely resembling real observations, we generate simulated images that are similar to the UVIT-N242W observations of goods south \citep{2024UVIT_Goods_south}. We randomly select sources in the \cite{2024UVIT_Goods_south} NUV catalog and use their magnitudes and half-light radius to generate 8 images with 500 sources in a square patch of $125\times 125$ arcsecond$^2$ ($300\times 300$ pixels). 
All sources were modeled using the 2D sersic model using astropy \citep{Astropy5}, with randomly selected position, Position angle, ellipticity, and sersic index between 0.5 to 5. These generated source fields were convolved with the model NUV PSF. 
To simulate observation images with these source fields, we add a flat NUV background ($\Tilde{B}\approx 1.16\times 10^{-4}$ cps/pixel corresponding to $\sim 27.7$ mag/arcsecond$^2$) from \cite{2024UVIT_Goods_south}, then this observation field ($\Omega$) which is in units of cps per pixel, is multiplied by an observation time ($t_{exp}$) to serve as the mean value for a Poisson random variable generator ($PoissonRV$) to make a full integrated time observation image ($I_{full}$). 
This $I_{full}$ is divided by $t_{exp}$ to create a cps per pixel observation image ($I$) as shown in equation \ref{eqn:Generated_UVIT_Image}.

\begin{align}
    I_{full}[i,j] &= PoissonRV(\Omega[i,j]\times t_{exp}) \nonumber \\
    I[i,j] &=\frac{I_{full}[i,j]}{t_{exp}} \label{eqn:Generated_UVIT_Image}
\end{align}

\begin{figure*}[ht]
\centering
    \includegraphics[width=1\textwidth]{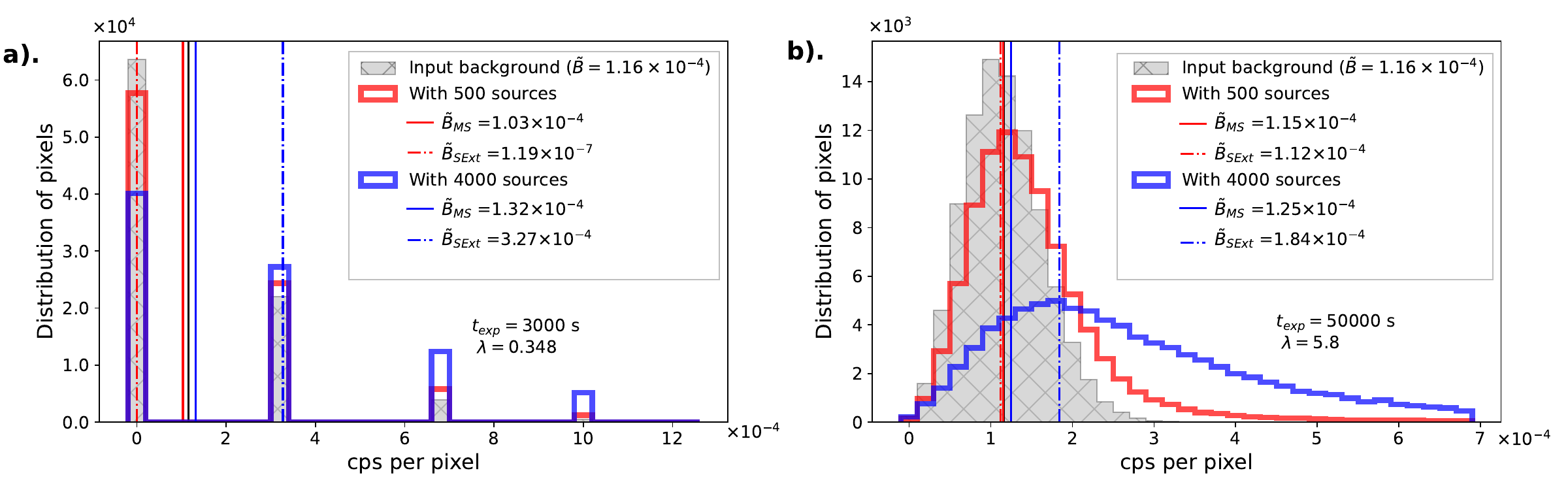}
    \caption{\textbf{(a)} The pixel histograms of simulated images with $t_{exp}=3000$ seconds (figure \ref{fig:fake_uvit_samp} b and e). \textbf{(b)}The pixel histograms of simulated images with $t_{exp}=50000$ seconds (figure \ref{fig:fake_uvit_samp} c and f). Poisson distributions of background are shown in grey with the input mean background ($\Tilde{B}=1.16\times 10^{-4}$ cps per pixel, shown by a black line) for both exposure times. The pixel histograms of images with 500 sources + background and 4000 sources + background are shown with red and blue color respectively. The recovery values of mean background with Minima Sampling routine and Sextractor are shown with solid and dot-dashed lines with same color as the histogram of the respective image }
    \label{fig:fake_obs_comp_hist}
\end{figure*}

To generate a very crowded field of $4000$ sources ($\sim$ 500 sources per square arcminute), we added all of the 8 generated source fields and repeated the process described above. 
Simulated UVIT observation images for $t_{exp}=$ 3000 and 50000 seconds, which correspond to background $\lambda=$ 0.348 and 5.8, hence a background rms of $1.9\times 10^{-4}$ and $4.8\times 10^{-5}$ cps per pixel respectively,
 are shown in figure \ref{fig:fake_uvit_samp} along with the source fields.

\begin{table*}[ht]
\hspace{-1.2cm}\resizebox{\textwidth}{!}{%
\begin{tabular}{cccccccc}
\hline
\multicolumn{1}{l}{Sources} & \multicolumn{1}{l}{$\Tilde{B}$(cps/pixel)} & \multicolumn{1}{l}{$t_{exp}$(s)} & \multicolumn{1}{l}{$\lambda=\Tilde{B}\times t_{exp}$} & \multicolumn{1}{l}{$\Tilde{B}_{MS}$ (cps/pixel)} & \multicolumn{1}{l}{Deviation in \%} & \multicolumn{1}{l}{$\Tilde{B}_{SExt}$ (cps/pixel)} & \multicolumn{1}{l}{Deviation in \%} \\ \hline
\multirow{2}{*}{500 }  & \multirow{2}{*}{$1.16\times 10^{-4}$}                  & 3000                            & 0.348                           & $1.05\times10^{-4}$                    & 9.32                                 & $1.19\times10^{-7}$                     & 99.89                                \\ 
                              &                                            & 50000                           & 5.8                             & 1$.15\times10^{-4}$                   & 0.65                                 & $1.12\times10^{-4}$                     & 3.1                                 \\ \hline
\multirow{2}{*}{4000 } & \multirow{2}{*}{$1.16\times10^{-4}$ }                  & 3000                            & 0.348                           & $1.32\times10^{-4}$                    & 13.79                                & $3.27\times10^{-4}$                     & 181.9                               \\ 
                              &                                            & 50000                           & 5.8                             & $1.25\times10^{-4}$                    & 9.84                                 & $1.84\times10^{-4}$                     & 58.27                                \\ \hline
\end{tabular}%
}
\caption{
Background recovery comparison in the UVIT-like Images shown in figure \ref{fig:fake_uvit_samp}. The detailed histograms of the images with the background recovery values is shown in Figure \ref{fig:fake_obs_comp_hist}} 
\label{table:ms_recovery_table2_UVIT}
\end{table*}

We ran the Minima Sampling routine and SExtractor background routine simultaneously on the simulated images and recovered mean background values from both methods ($\Tilde{B}_{MS}$ and $\Tilde{B}_{SExt}$ respectively, which are presented in table \ref{table:ms_recovery_table2_UVIT}.

In case of low lambda background ($\lambda = 0.348$ or $t_{exp}=3000$ s), we find that in the uncrowded field with 500 sources, the median count is 0 and for the crowded field with 4000 sources, the median count of the pixel distribution is $1/t_{exp}$ or $3.33\times 10^{-4}$ cps per pixel and the mean background values recovered by SExtractor ($\Tilde{B}_{SExt}$) are $1.19 \times 10^{-7}$ (35.5 mag/arcsec$^2$) and $3.27 \times 10^{-4}$ cps/pixel (26.5 mag/arcsec$^2$), which are near the median count values (as seen in figure \ref{fig:fake_obs_comp_hist}a) and absurdly away from the input background with a fixed mean of $1.16\times 10^{-4}$ cps per pixel.
But the Minima Sampling routine is able to recover the mean background value ($\Tilde{B}_{MS}$) closer to the real mean (figure \ref{fig:fake_obs_comp_hist}a) with values $1.03\times 10^{-4}$ (27.83 mag/arcsec$^2$) and $1.32\times 10^{-4}$ (27.56 mag/arcsec$^2$), with a deviation which is lower compared to SExtractor. 

In the case of medium-high lambda background ($\lambda = 5.8$ or $t_{exp}=50000$ s), we see that the background distribution now resembles a Gaussian (figure \ref{fig:fake_obs_comp_hist}b) which is a range where 3$\sigma$ clipping algorithms of SExtractor's background routine works well as seen in the case of the uncrowded field(with 500 sources) as it recovers a background value of $1.12\times 10^{-4}$ cps/pixel (27.74 mag/arcsec$^2$) and minima sampling routine recovers a value  $1.15\times 10^{-4}$ cps/pixel (27.71 mag/arcsec$^2$)
But in the case of crowded field (with 4000 sources), we see that the mode, which is the converging point of $3\sigma$ clipping algorithms in general, has shifted rightwards,  which caused SExtractor to overestimate the background by $\approx 58\%$ (table \ref{table:ms_recovery_table2_UVIT}) with a value of $1.84\times 10^{-4}$ cps/pixel (27.2 mag/arcsec$^2$). 
However, this shift in mode affects the Minima sampling routine to a lesser degree, and it recovers $\Tilde{B}_{MS}=1.25\times10^{-4}$ cps/pixel (27.62 mag/arcsec$^2$), which is closer to the input mean background.
Overall from the results shown in table \ref{table:ms_recovery_table2_UVIT}, we find that the deviations in mean background recovery by Minima Sampling routine from the input background value are about an order of magnitude lower than the deviation from SExtractor's background routine, both in crowded and uncrowded, and both in low $\lambda$ and high $\lambda$ observations.

\subsection{Spatially Varying Background Map}

In large observation fields, the background may not be a constant value throughout the image, and it may be spatially varying. In such cases, we can use the minima sampling method to create a background map using the following routine extension.
\begin{enumerate}
    \item Identify all the minima regions in the image beforehand and mark the minima locations using the recipe in Section 3.1.
    \item Prepare a grid of appropriate size for creating the background and variance map. We suggest dividing the image into patch sizes of $>20$ times the PSF FWHM of an image.
    \item For each patch, we can look for the minima locations within each patch for background estimation.
    \item In case a bigger source is present or a large number of sources cover an entire patch resulting in very little or no minima inside the patch, we can look for an appropriate number of minima locations near it for performing a reliable background estimation at that location.
    \item For a continuous and smooth background map, we could interpolate between the patches using linear or 2d spline interpolation functions.
    \item Ideally, for purely Poissonian background, the rms or standard deviation map would follow the relation 
\end{enumerate}
$$I_{\sigma}(x,y)=\frac{\sqrt{I_{exp}(x,y)\times I_{background}(x,y)}}{I_{exp}(x,y)}$$

\noindent Where $I_{\sigma}$, $I_{exp}$ and $I_{background}$ are the rms, exposure and background maps of the image. But in case of the presence of other significant additional noise layers on the image, we can use the same recipe for background maps and calculate standard deviation or rms within the minima pixels to create a reliable standard deviation/rms map. 

To select an appropriate patch size for dividing the full field of view (FOV) into a grid for background mapping, we need to ensure a good sample size of pixels for a reliable background estimate. If the initial smoothing kernel ($G_1$) is chosen such that its FWHM is $10 \times$ the PSF FWHM of the field, the resulting image typically yields around one minimum per $10 \times 10$ PSF-FWHM$^2$ area. When using a smaller kernel ($G_2$) with a scale similar to the PSF, the number of minima detected typically increases by a factor of 4–5. Setting a sampling box size of $7 \times 7$ pixels allows for approximately 200 pixels within a $10 \times 10$ PSF-FWHM$^2$ patch. For a statistically reliable background estimate, sampling around 1000 pixels is preferable, suggesting a larger patch size of roughly $20 \times 20$ PSF-FWHM$^2$. These approximations help account for sky variability on the scale of the chosen patch size, as using a patch size that’s too small may lead to less robust background estimates.

We have already applied the minima routine to calculate the background and rms maps for the AUDF GOODS South field; see the Background maps shown in figure 9 of \cite{2024UVIT_Goods_south}.
Our Python code, following the routine, was able to calculate the Background and RMS map of 236 arcminutes$^2$ (1897$\times$ 2580 pixel$^2$) observation data in about 30 seconds in a system with a 2.5GHz processor on a single core, which is fast considering its applicability on such observation fields. 

\section{Discussion and conclusions}
\label{sec:discuss}

In this section, we discuss the bias and limitations of the proposed minima methods and its advantage over the current standard 3-sigma clipping methods for background estimation. We also discuss a basic method to employ the minima method to make background maps for wide-field observations.

\smallskip
\noindent \textbf{Bias and Limitations:}\\
The minima sampling methods from both section \ref{section:method1} and section \ref{sec:crowded} have sampling bias as we only sample pixels at minima or around minima locations, which causes the mean of our sample to shift leftwards or in other words the mean of selected sample would always be less than the mean of actual background. 
The correction of this bias is an extra term (as $K \sigma_{min}$ and $F\sigma$ of equation \ref{mu_recover} and \ref{smoth_lambdarec}), which should be added to the minima mean ($\mu_{min}$) for correct estimation of the background, but as we move to images with sources, things become more complicated. 
The general distribution of sources, however, is an unknown for any given observation patch, and this distribution of sources and the PSF (which dictates the spread of an individual source) could have a significant impact on the background estimation, which is added as an additional term $S_c$ in equation \ref{smoth_murec_test_Sc}. 
The source contamination term $S_c$ causes a subtle overestimation in lower source density fields and a large overestimation in highly crowded fields.  In general, this type of overestimation of background is accepted as confusion noise due to unresolved sources, and to dampen the impact of this bump of background due to undetected sources in such crowded fields, we may choose to ignore the initial bias correction in case of large sampling box sizes ($\sim 0.1\sigma$ for $b=11$ and $\sim 0.01\sigma$ for $b=25$), as the source contamination from such large sampling boxes will almost always higher than the correction factor in comparison.

While the minima sampling method performs much better at background estimation than the standard Gaussian approximation based on 3 sigma clipping methods for high noise regimes (where $\lambda<1$), we still cannot be confident about the regime where $\lambda\lesssim 0.1$ as the recovery plot in figure \ref{fig:graphs} shows a deviating feature when $\lambda_{input}\sim 0.1$. It is advisable in such cases to use wider sampling boxes to overcome the problem of having a very low photon count in images.

Another limitation of the minima sampling routine arises with diffuse objects of large angular sizes (on the arcminute to degree scale), as these objects may be misinterpreted as part of the background. The method was developed and tested with smaller objects (on the arcseconds scale), and while it reliably finds empty regions for background sampling, these regions are identified locally. A very large diffuse object could be mistaken for an area of uniformly elevated background, leading to potential minima detections within the object itself. To use the minima sampling routine with such objects, we suggest that the initial smoothing kernel ($G_1$) should be large enough to smooth the entire object, ensuring no initial minima are detected in the object itself. Further tests are required to test the effectiveness of the minima routine on such objects

\medskip

\noindent \textbf{Advantages over Traditional Background Estimators:}\\
A direct noticeable advantage of the minima sampling method over the traditional 3-sigma clipping-based background estimation methods is its approach to a 2-dimensional search for minima in an image which preserves and utilizes the 2-dimensional distribution of pixel values, rather than the sigma clipping method where all the pixels are collapsed into a 1-D distribution to find the median/mode background values. 
In these traditional methods, a very faint source with pixel values not brighter than a few $\sigma$ would contribute towards the background estimation, and a crowded field of such faint sources will force the mode to move rightwards and force overestimation of the background. While crowding still causes overestimation of background with minima sampling, the deviation from true background is very low compared to the traditional methods.

Looking for minima away from bright sources increases accuracy at finding background-dominated regions in the low $\lambda$ domain of observations (evident in figure \ref{fig:sourc_cont_bar}). This feature allows us to recover the mean background in an image closer to the true background in observation images.

\medskip 
\noindent\textbf{Suitable application domain:}\\
While detecting sources, background calculation rarely matters for detecting very bright sources (compared to background). The low surface brightness sources with pixel values $\lesssim\mathcal{O}(\sigma)$ are the most affected by even slightly overestimating background and noise values. In the case of crowded fields, we have shown that the traditional background methods tend to overestimate the background, which would then virtually suppress the signal-to-noise values of faint sources, making their detection more questionable than it should have been. 

We propose using the minima method for background estimation in all astronomy images where the suspected density of faint sources is higher. Using the minima method, we avoid much of the source light and crowded regions to sample uncontaminated background. Using minima sampling would facilitate detection in higher numbers of faint sources near the 3 SNR limit. Our method could be applied to wide-field deep sky surveys such as Euclid \citep{EuclidCollab2022}, DESI imaging surveys \citep{Deyetal2019}, LSST \citep{2019LSST}.

\medskip
Our primary conclusions from this work are the following:
\begin{itemize}
    \item We have developed a Background sampling routine in Astronomy images, which targets the source-free/background-dominated regions in 2-D astronomy data using minima to sample the pixels that are least contaminated by source fluxes. The source contamination fraction is found to be 1-2 orders of magnitude lower for the Minima sampling routine than the traditional Background estimator (e.g., SExtractor).
    \item In the case of a low-photon counting image ($\lambda=0.348$) with low source density, a traditional background estimator (e.g., SExtractor) underestimates the background by several orders of magnitude from the true background, whereas our Minima sampling method recovers the background within 10\% of the true background.
    \item In the case of a low photon counting image and crowded field, the traditional method overestimates the background by about $\sim 3 \times$ (or $\sim 200\%$), while the Minima sampling method recovered the background within 14\%.
    \item In the case of medium-high photon count images($\lambda=5.8$), where the background distribution resembles a bell curve, In the case of a crowded field, the traditional background estimator overestimates the background by 60\% whereas, Minima sampling overestimates the background by only $\sim 10\%$. \item Both methods perform well in the case of an uncrowded field, where the traditional method underestimates the background by $\sim 3\%$ and Minima sampling underestimates the background by a mere $0.65\%$.
\end{itemize}

\software{astropy \citep{Astropy5}, scipy\citep{2020SciPy-NMeth}, 
          Source Extractor \citep{SExtractor}
          }

\label{lastpage}

\bibliography{Minima_dist_arxiv_sub}{}
\bibliographystyle{aasjournal}

\appendix
\section{The mean shift coefficient and sigma ratio for Gaussian random fields }
\label{section:Appendix_A}

\begin{figure*}[ht]
\centering
    \includegraphics[width=1\textwidth]{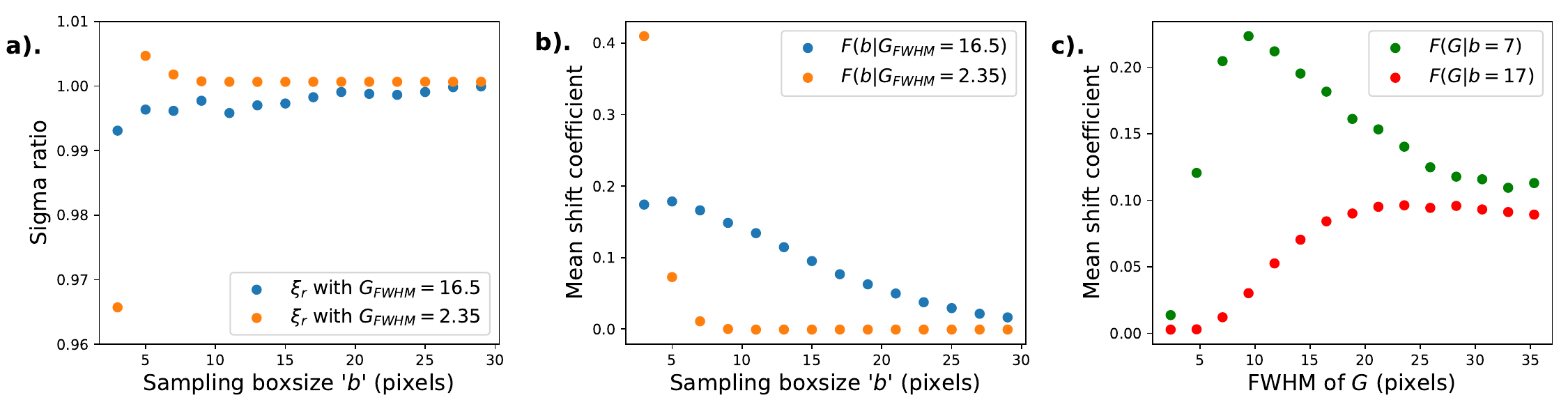}
    \caption{\textbf{(a)}  Sigma ratio($\xi_r$) vs sampling boxsize($b$) with fixed FWHM of $G=16.5$ pixels (cyan-blue) and $G=2.35$ pixels (orange). \textbf{(b)} calculated mean shift coefficient vs. sampling box size with fixed FWHM of $G=16.5$ pixels (cyan-blue) and $G=2.35$ pixels (orange). \textbf{(c)} mean shift coefficient's dependence on the  smoothing kernel ($G$) while keeping $b$ fixed at $7$ pixels (green) and $17$ (red).}
    \label{fig:mean_coefficient_1}
\end{figure*}

To estimate background using the smoothed image minima in section \ref{sec:crowded}, we introduce the term Mean shift coefficient $\bigl(F(G,b)\bigr)$, which relates the difference between the mean of minima distribution ($\mu_{min}$) and mean of parent distribution ($\mu$) to the standard deviation of minima distribution ($\sigma_{min}$), as shown in equation \ref{smoth_lambdarec}. $F(G,b)$ depends on the smoothing kernel $(G)$ and sampling boxsize $(b)$.
We calculate $F(G,b)$ for different $G$ and $b$ in simulated Gaussian fields of size $1000\times 1000$ pixels to get a general idea of $F(G,b)$'s dependence on its parameters. 
 We define sigma ratio $(\xi_r)$ as 
\begin{equation}
    \xi_r=\frac{\sigma_{min}}{\sigma} 
\end{equation}   
where $\sigma$ is the standard deviation of the parent distribution of background pixels. We find that $\sigma_{min}\approx \sigma$ within $0.5\%$ relative error ( or $|1-\xi_r|<0.005$)  for $b\geq 7$ pixels for two different $G$ (figure \ref{fig:mean_coefficient_1}a).
The dependence of $F(G,b)$ on $b$ in the case of smoothing is also shown for two different G in figure \ref{fig:mean_coefficient_1}b.
The dependence of $F(G,b)$ on $G$ is shown in figure \ref{fig:mean_coefficient_1}c, where G is a 2d gaussian shaped smoothing kernel with FWHM ranging from 2.35 to 35.32 for two different values of $b$.

For the case of Minima Sampling routine developed in section \ref{sec:crowded}, we make use of two kernels as shown in figure \ref{fig:Minima2Flowchart}. The corresponding mean shift equation then becomes
\begin{equation}
    \mu - \mu_{min} = F(G_1,G_2,b)\sigma_{min}
    \label{smoth_lambdarec_appn}
\end{equation}

\begin{figure*}[ht]
\centering
    \includegraphics[width=1\textwidth]{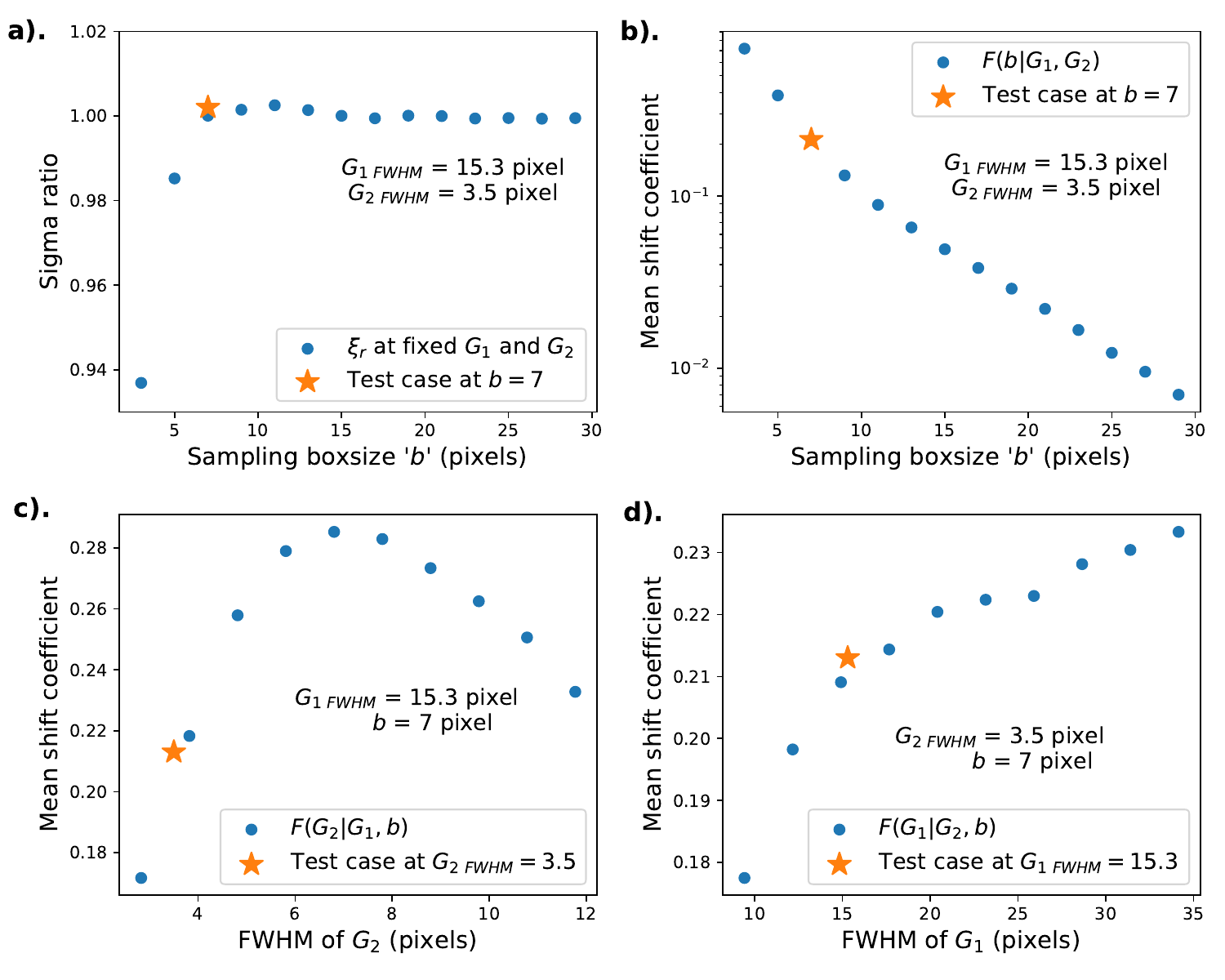}
    \caption{\textbf{(a)}  Sigma ratio($\sigma_{min}/\sigma$) vs sampling boxsize($b$) with fixed FWHM of $G_1=15.3$ and $G_2=3.5$ pixels. \textbf{(b)} calculated mean shift coefficient vs sampling boxsize with fixed FWHM of $G_1=15.3$ and $G_2=3.5$ pixels.\textbf{(c)} mean shift coefficient's dependence on the final smoothing kernel($G_2$) while keeping FWHM of $G_1 = 15.3$ and $b=7$. \textbf{(d)} mean shift coefficient's dependence on the initial smoothing kernel($G_1$) while keeping FWHM of $G_2 = 3.5$ and $b=7$.}
    \label{fig:mean_coefficient}
\end{figure*}

The relation of the sigma ratio to sampling size for minima sampling routine is shown in figure \ref{fig:mean_coefficient}(a) for $3\leq b \leq 29$ pixels in a test performed on generated Gaussian fields. Following this, We can rewrite the mean shift  coefficient in the equation \ref{smoth_lambdarec_appn} as
\begin{equation}
    F(G_1,G_2,b)=\frac{\mu - \mu_{min}}{\sigma_{min}}\approx \frac{\mu - \mu_{min}}{\sigma}
\end{equation}

This mean shift coefficient was tested for its dependence upon $b$, $G_2$, and $G_1$ in generated Gaussian fields. Where to test the dependence on one of the parameters, we fix the other two. We calculate empirical values of $F(G_1, G_2,b)$ for a range of $3\leq b \leq 29$, $2.8\leq G_2 \leq 11.7$ and $9.4\leq G_1 \leq 34$. The fixed values for the parameters were kept the same as our test case presented in section \ref{sec:crowded}, which are $b=7$, $G_2 = 3.5$ and $G_1 = 15.3$ pixels. The results of these tests are shown in figure \ref{fig:mean_coefficient}.

The dependence of $F(G,b)$ and $F(G_1, G_2,b)$ on $b$ is inverse in all the tested cases, suggesting that the mean shift ($\mu-\mu_{min}$) is lower at higher sampling box sizes, but as discussed in the text, we cannot choose arbitrary $b$ size, fearing source contamination in the sampled pixels, but a very low sampling box size will be responsible for high mean shift, which would require correction for background sampling. The scale length of this inverse relation is proportional to the smoothing kernels used in both $F(G,b)$ and $F(G_1,G_2,b)$.
From figure \ref{fig:mean_coefficient_1}c, we see that $F(G,b)$ rises with $G_{FWHM}$ till $G_{FWHM}\lesssim b$, The peak value of $F(G,b)$ occurs at a value of $G_{FWHM} \propto b$, and then it declines with $G_{FWHM}$. This behavior is again repeated by the mean shift coefficient of the minima sampling method $\bigl(F(G_1,G_2,b)\bigr)$ for $G_2$ (figure \ref{fig:mean_coefficient}c). The dependence of $F(G_1,G_2,b)$ on $G_1$ is positive but weaker than $G_2$ and $b$ as shown by figure \ref{fig:mean_coefficient}.

\end{document}